\documentclass[prl,aps,showpacs,twocolumn]{revtex4}
\usepackage{epsfig}
\usepackage{amsmath}

\begin{document}

\title{Berry Phase of Nonlinear Correction}
\author{J. Liu$^{1,2}$ and L.B. Fu$^{2}$}
\affiliation{1. College of Physics and Information, Beijing Institute of Technology,
Beijing 100081, China \\
2. Institute of Applied Physics and Computational Mathematics, P.O. Box
8009, Beijing 100088, China}

\begin{abstract}
We investigate the geometric phase or Berry phase of adiabatic quantum
evolution in the Bose-Einstein condensate (BEC) systems governed by
nonlinear Gross-Pitaevskii(GP) equations. We study how this phase is
modified by the nonlinearity and find that the Bogoliubov fluctuations
around the eigenstates are accumulated during the nonlinear adiabatic
evolution and contribute a finite phase of geometric nature. A two-mode BEC
model is used to illustrate our theory. Our theory is applicable to other
nonlinear systems such as paraxial wave equation for nonlinear optics and
Ginzburg-Landau equations for complex order parameters in condensed-matter
physics.
\end{abstract}

\author{}
\maketitle

Adiabatic theory, as a fundamental issue of quantum mechanics, involves two
aspects: i) when the Hamiltonian changes slowly compared to the level
spacings, an initial nondegenerate eigenstate remains to be an instantaneous
eigenstate\cite{theorem1}; ii) the phase acquired by the eigenstate is the
sum of the time integral of the eigenenergy (dynamical phase) and a quantity
independent of the time duration and related to the geometric property of
the closed path in parameter space (geometric phase or Berry phase)\cite%
{theorem2,liu2}. The adiabatic theory has been playing a crucial role in
preparation and control of quantum states\cite{application}. Recently, the
Berry phase and related geometric phases\cite{fu3,fu4} has received renewed
interest due to its important use in implementation of quantum computing
gates\cite{renew} and applications in condense matter physics\cite{fu6}.

For a nonlinear quantum system, such as that described by the nonlinear Schr%
\"{o}dinger equations, how the adiabatic theory gets modified. Nonlinear
quantum systems have become increasingly important in physics. They often
arise in the mean field treatment of many-body quantum systems, such as
Bose-Einstein condensates (BECs) of dilute atomic gases\cite{bec}.
Recently, extending the first aspect of adiabatic theory to the nonlinear
systems, i.e., investigating the adiabatic condition and adiabaticity for
the nonlinear quantum evolution has been done \cite{liu,pu}, interestingly
it was found that the adiabaticity of an eigenstate only requires that the
control parameters vary slowly with respect to the Bogoliubov excitation
frequencies and has nothing to do with the level spacings. Nevertheless,
Berry phase issue in such nonlinear system is far from well understand,
while some superficial observations suggest that Berry's formula without any
correction is applicable to nonlinear system suppose the system is invariant
under gauge transformation of the first kind \cite{chiao}.

In this letter, we make a thorough analysis on geometric phase associated
with the adiabatic evolution of an eigenstate, and strikingly we find that
the Berry phase is dramatically modified by the nonlinearity. This finding
is completely contrary to previous superficial observations. The underlying
mechanism has been revealed: for a nonlinear system because the Hamiltonian
is a functional of the instantaneous wavefunctions, the Bogoliubov
fluctuations around the eigenstate caused by the slow change of the system
are allowed to feedback to the Hamiltonian. They are accumulated during an
adiabatic evolution and eventually contribute a finite phase of geometric
nature. A two-mode BEC model is used to illustrate our theory.

Let us consider the BEC trapped in a potential $V(\mathbf{R};r)$, the
evolution of its wavefunction is governed by following nonlinear
Gross-Pitaevskii(GP) equation ($\hbar =m=1$)\cite{GP},
\begin{equation}
i\frac{\partial \psi }{\partial t}=H_{0}\psi +g|\psi |^{2}\psi ,  \label{sc}
\end{equation}%
with $H_{0}=-\frac{1}{2}\bigtriangledown ^{2}+V(\mathbf{R};r),$ where $%
\mathbf{R}$ is the parameter vector, $g$ is the nonlinear parameter
representing the interaction between the coherent atoms. The total energy of
the system $\mathit{E}_{T}=\int drE(\psi ^{\ast },\psi )$, where the energy
density $E(\psi ,\psi ^{\ast })=\psi ^{\ast }H_{0}\psi +\frac{1}{2}g|\psi
|^{4}.$ The above system is invariant under gauge transforms of the first
kind, $\psi (r,t)\rightarrow \exp (i\eta )\psi (r,t)$ with constant $\eta .$
The gauge symmetry implies that the total atom number is conserved, i.e., $%
\int dr|\psi |^{2}=1$.

Let $\lambda $ be the overall phase of the wave function; we may take it to
be the phase of the wavefunction at a fixed position $r_{0}$, for example, $%
\lambda =-\arg (\psi (r_{0},t)).$ We split off this overall phase by writing
$\psi =e^{-i\lambda }\phi ,$ then $\phi $ belongs to the so called
projective Hilbert space. From (\ref{sc}) we obtain
\begin{equation}
\frac{d\lambda }{dt}=-i\langle \phi |\frac{\partial }{\partial t}|\phi
\rangle +\int drE(\phi ^{\ast },\phi )+\frac{g}{2}\langle \phi |\phi ^{\ast
}\phi |\phi \rangle .  \label{da}
\end{equation}

The eigenequation of the system is
\begin{equation}
H_{0}\overline{\psi }+g|\overline{\psi }|^{2}\overline{\psi }=\mu \overline{%
\psi }.
\end{equation}
where $\overline{\psi }$ is the eigenfunction and $\mu $ is the eigenvalue
(or chemical potential).

Now we consider the parameter vector $\mathbf{R}$ varies slowly in time, and
introduce the dimensionless adiabatic parameter of $\varepsilon \sim |\frac{d%
\mathbf{R}}{dt}|$ as the measure how slow the parameters change. The
adiabatic parameter tends to zero, i.e., $\varepsilon \rightarrow 0$,
indicating the adiabatic limit.

Consequently, the expression of the total phase can be expanded in a
perturbation series in the adiabatic parameter, i.e.,
\begin{equation}
\frac{d\lambda }{dt}=\alpha _{0}(\varepsilon ^{0})+\alpha _{1}(\varepsilon
^{1})+o(\varepsilon ^{2}).
\end{equation}%
When the parameters move in a circuit, the eigenstates evolves for an
infinite long time duration in the adiabatic limit. The time integral of the
zero-order term gives so-called dynamic phase because it is closely related
to the temporal process of the evolution. The time integral of the
first-order term gives an additional contribution to the overall phase,
which will be shown later is of a geometric nature, that is, only depends on
the geometry of the close path in the parameter space. The contribution of
the higher-order term vanishes in the adiabatic limit.

In the quantum evolution with slowly-changing parameters, we assume $\phi =%
\overline{\phi }(\mathbf{R})+\delta \phi (\mathbf{R}),$ where $\overline{%
\phi }(\mathbf{R})$ is the wavefunction of the instantaneous eigenstate
corresponding to the local minimum energy. $\delta \phi $ denotes the
secular part of the Bogoliubov fluctuations induced by the system's slow
change while the rapid oscillations in the fluctuations are ignored because
they vanish after a long-term average. $\delta \phi $ depends on the
adiabatic parameter and is of order $\varepsilon ,$ then from Eq.(\ref{da})
and with the help of relation Eq.(3), we have the explicit expressions as
follows,
\begin{equation}
\alpha _{0}(\varepsilon ^{0})=\mu (\mathbf{R}),
\end{equation}%
\begin{equation}
\alpha _{1}(\varepsilon ^{1})=-i\langle \overline{\phi }|\frac{\partial }{%
\partial t}|\overline{\phi }\rangle +g\langle (%
\begin{array}{cc}
\overline{\phi }^{2}\overline{\phi }^{\ast }, & \overline{\phi }^{\ast 2}%
\overline{\phi }%
\end{array}%
)|(\delta \phi ,\delta \phi ^{\ast })\rangle ,  \label{as}
\end{equation}%
where we denote $\langle a|b\rangle =\int dr(a^{\ast }b),\langle
(a,b)|(c,d)\rangle =\int dr(a^{\ast }c+b^{\ast }d),$and the second order
term like $\frac{\partial \delta \phi }{\partial t}$has been ignored.

From the above expressions we see that the dynamical phase has been modified
to be the time integral of the chemical potential rather than the energy.
This is because the instantaneous eigenstates are feedback to the
Hamiltonian. More interestingly, the first-order term, i.e., the Berry phase
term has been modified due to the feedback of the Bogoliubov fluctuations to
the Hamiltonian. To evaluate it qualitatively and express the modified
geometric phase explicitly, let us introduce \ a set of orthogonal basis $%
|k\rangle $ and the variable $\psi _{j}$ is the j-th component, i.e., $\psi
_{j}=$ $\langle j|\psi \rangle .$ Without losing generality, the projective
Hilbert space is set to be of a specific gauge that the phase of the N-th
component is zero. In the projective Hilbert space, \ the new variables ($%
n_{j}$,$\theta _{j})$ are introduced through $\phi _{j}=\sqrt{n_{j}}%
e^{i\theta _{j}}.$ Substituting the expression of $\psi _{j}=\sqrt{n_{j}}%
e^{i\theta _{j}}e^{-i\int^{t}\beta dt}$ into the GP equation, and separating
real and imaginary parts, we have following differential equations for the
density $n_{j}$ and phase $\theta _{j},$ respectively,

\begin{equation}
\frac{dn_{j}}{dt}=f_{j}\text{ , }\frac{d\theta _{j}}{dt}=h_{j},\text{ }%
j=1,2,...N-1
\end{equation}%
where%
\begin{eqnarray}
f_{j}\text{ } =2\sum_{k=1}^{N}n\sqrt{n_{j}n_{k}}\text{Im}\left[ C_{jk}(%
\mathbf{R)}e^{i(\theta _{k}-\theta _{j})}\right]  \notag \\
+2g\sum_{k,l,m=1}^{N}\sqrt{n_{j}n_{k}n_{l}n_{m}}\text{Im}[D_{j,k,l,m}e^{i(%
\theta _{l}+\theta _{m}-\theta _{k}-\theta _{j})}], \\
h_{j} =-\sum_{k=1}^{N}\sqrt{\frac{n_{k}}{n_{j}}}\text{Re}\left[ C_{jk}(%
\mathbf{R)}e^{i(\theta _{k}-\theta _{j})}\right]  \notag \\
-g\sum_{k,l,m=1}^{N}\sqrt{\frac{n_{k}n_{l}n_{m}}{n_{j}}}\text{Re}%
[D_{j,k,l,m}e^{i(\theta _{l}+\theta _{m}-\theta _{k}-\theta _{j})}]+\beta ,
\label{gi} \\
\text{ }\beta =\sum_{k=1}^{N}\sqrt{\frac{n_{k}}{n_{N}}}\text{Re}\left[
C_{Nk}(\mathbf{R)}e^{i(\theta _{k}-\theta _{N})}\right] &  \notag \\
+g\sum_{k,l,m=1}^{N}\sqrt{\frac{n_{k}n_{l}n_{m}}{n_{N}}}\text{Re}%
[D_{N,k,l,m}e^{i(\theta _{l}+\theta _{m}-\theta _{k}-\theta _{N})}].
\end{eqnarray}%
with $C_{jk}(\mathbf{R)}=\langle j|H_{0}(\mathbf{R)}|k\rangle
,D_{j,k,l,m}=\langle j|\langle k|l\rangle |m\rangle .$

The last equation is from the fact that in the projective Hilbert space the
phase of wavefunction of the N-th component is set to be zero. The norm
conversation condition $n_{N}=1-\sum_{k=1}^{N-1}n_{k}$ could be used to
remove the variable $n_{N}$ in the above equations. In the representation of
new variables, ($\overline{n}_{j}$,$\overline{\theta }_{j})$ satisfy
equations of the equilibrium state, i.e., $\left( \frac{\partial n_{j}}{%
\partial t},\frac{\partial \theta _{j}}{\partial t}\right) |_{(\overline{n}%
_{j},\overline{\theta }_{j})}=0.$ ($\overline{n}_{j}$,$\overline{\theta }%
_{j})$ are function of the parameter $\mathbf{R}$ corresponding to the
eigenstates of GP equation. Let us make perturbation expansion around the
eigenstate with $n_{j}=\overline{n}_{j}(\mathbf{R})+\delta n_{j}(\mathbf{R}%
),\theta _{j}=\overline{\theta }_{j}(\mathbf{R})+\delta \theta _{j}(\mathbf{R%
}).$ Here, $\overline{\phi }_{j}(\mathbf{R})=\sqrt{\overline{n}_{j}(\mathbf{R%
})}e^{i\overline{\theta }_{j}(\mathbf{R})},$ ($\delta n_{j}(\mathbf{R}%
),\delta \theta _{j}(\mathbf{R}))$ are the fluctuations depending on the
adiabatic parameter and of order $\varepsilon $. Then inserting the above
expansion into the equations (7) and ignoring the higher order terms such as
$\frac{\partial \delta n_{j}}{\partial t},\frac{\partial \delta \theta _{j}}{%
\partial t},$ with denoting $\upsilon =(n_{1},\theta _{1};...;n_{N-1},\theta
_{N-1})$ we obtain that%
\begin{equation}
\frac{d\overline{\upsilon }}{d\mathbf{R}}\frac{d\mathbf{R}}{dt}=\mathcal{L}%
\delta \upsilon ,
\end{equation}%
where the matrix takes the form,
\begin{equation}
\mathcal{L}=\left\{ \mathcal{L}_{jk}\right\} _{_{(N-1,N-1)}},\mathcal{L}%
_{jk}=\left(
\begin{array}{cc}
\frac{\partial f_{j}}{\partial n_{k}} & \frac{\partial f_{j}}{\partial
\theta _{k}} \\
\frac{\partial h_{j}}{\partial n_{k}} & \frac{\partial h_{j}}{\partial
\theta _{k}}%
\end{array}%
\right) _{\upsilon =\overline{\upsilon }}.
\end{equation}%
then, inversely we have,%
\begin{equation}
\delta \upsilon =\mathcal{L}^{-1}\frac{d\overline{\upsilon }}{d\mathbf{R}}%
\frac{d\mathbf{R}}{dt}.  \label{dv}
\end{equation}

The differential relation between the new variables and old ones take the
form
\begin{equation}
\left(
\begin{array}{c}
\delta \phi _{j} \\
\delta \phi _{j}^{\ast }%
\end{array}%
\right) =\Pi _{j}\left(
\begin{array}{c}
\delta n_{j} \\
\delta \theta _{j}%
\end{array}%
\right) .  \label{pj}
\end{equation}%
in which
\begin{equation}
\Pi _{j}=\left(
\begin{array}{cc}
\frac{1}{2}\overline{n}_{j}^{-1/2}e^{i\overline{\theta }_{j}} & i\sqrt{%
\overline{n}_{j}}e^{i\overline{\theta }_{j}} \\
\frac{1}{2}\overline{n}_{j}^{-1/2}e^{-i\overline{\theta }_{j}} & -i\sqrt{%
\overline{n}_{j}}e^{-i\overline{\theta }_{j}}%
\end{array}%
\right) .
\end{equation}%
Substituting (\ref{dv}) and (\ref{pj}) into (\ref{as}), we finally obtain
the explicit expression of adiabatic geometric phase that contains two
terms,
\begin{equation}
\gamma _{g}=\gamma _{B}+\gamma _{NL}  \label{ggg}
\end{equation}%
where the first term is the usual Berry phase formula,

\begin{equation}
\gamma _{B}=-i\oint \langle \overline{\phi }|\nabla _{\mathbf{R}}|\overline{%
\phi }\rangle d\mathbf{R=}\oint \sum_{j=1}^{N-1}\overline{n}_{j}\frac{%
\partial \overline{\theta }_{j}}{\partial \mathbf{R}}d\mathbf{R}.  \label{gb}
\end{equation}%
and additional term is from the nonlinearity, taking the form,
\begin{equation}
\gamma _{NL}=g\oint \left\langle \Lambda |\Pi \mathcal{\circ L}^{-1}|\frac{d%
\overline{\upsilon }}{d\mathbf{R}}\right\rangle d\mathbf{R}.  \label{lf}
\end{equation}%
Here, $\Lambda =((\overline{n}_{1}+\sum\limits_{j=1}^{N-1}\overline{n}_{j}-1)%
\overline{n}_{1}^{1/2}e^{i\overline{\theta }_{1}},(\overline{n}%
_{1}+\sum\limits_{j=1}^{N-1}\overline{n}_{j}-1)\overline{n}_{1}^{1/2}e^{-i%
\overline{\theta }_{1}},$ $\cdots $ $(\overline{n}_{N-1}+\sum%
\limits_{j=1}^{N-1}\overline{n}_{j}-1)\overline{n}_{N-1}^{1/2}e^{i\overline{%
\theta }_{N-1}},(\overline{n}_{N-1}+\sum\limits_{j=1}^{N-1}\overline{n}%
_{j}-1)\overline{n}_{N-1}^{1/2}e^{-i\overline{\theta }_{N-1}}),$ $\frac{d%
\overline{\upsilon }}{d\mathbf{R}}=(\frac{d\overline{n}_{1}}{d\mathbf{R}},%
\frac{d\overline{\theta }_{1}}{d\mathbf{R}},\cdots \frac{d\overline{n}_{N-1}%
}{d\mathbf{R}},\frac{d\overline{\theta }_{N-1}}{d\mathbf{R}})^{T}$, and
diagonal matrix $\Pi =\text{diag}(\Pi _{1},\Pi _{2},...\Pi _{N-1}).$Notice
that to simplify the expression of $\Lambda ,$ we use the approximation that
the overlap intergral $D_{j,k,l,m}\simeq 0$ when the subscripts are not\ all
identical.

Both $\gamma _{B}$ and $\gamma _{NL}$ have the geometric property of
parameter space. The novel second term indicates that, the Bogoliubov
fluctuations induced by the slow change of the system that is negligible in
linear case, however, could be accumulated in the nonlinear adiabatic
evolution and contribute the finite phase of a geometric nature. This is
because the Hamiltonian in the nonlinear system contains the instantaneous
wavefunction and as a result the Bogoliubov fluctuations around the
eigenstate are allowed to feedback to the Hamiltonian.

As an illustration of our theoretical formulism, we consider the two-mode
BEC model as example, which is described by the following GP equation\cite%
{twomode},
\begin{equation}
i\frac{d}{dt}\left(
\begin{array}{c}
\Psi _{1} \\
\Psi _{2}%
\end{array}%
\right) =H(\Psi _{1},\Psi _{2})\left(
\begin{array}{c}
\Psi _{1} \\
\Psi _{2}%
\end{array}%
\right) ,\text{ }  \label{sss}
\end{equation}%
with%
\begin{equation}
H(\Psi _{1},\Psi _{2})=\left(
\begin{array}{cc}
Z|\Psi _{1}|^{2} & \frac{\rho }{2}e^{-i\varphi } \\
\frac{\rho }{2}e^{i\varphi } & Z|\Psi _{2}|^{2}%
\end{array}%
\right)
\end{equation}%
the $\mathbf{R}=(\rho ,\varphi ,Z)$ are parameters. For simplicity, we fix $%
\rho $ and $Z$ and change the parameter $\varphi $ from 0 to 2$\pi $
adiabatically.

The energy of the system $E(\Phi _{1},\Phi _{2})=\frac{Z}{2}(|\Psi
_{1}|^{4}+|\Psi _{2}|^{4})+\frac{\rho }{2}(e^{-i\varphi }\Psi _{1}^{\ast
}\Psi _{2}+e^{i\varphi }\Psi _{1}\Psi _{2}^{\ast }).$The above system is
invariant under gauge transforms of the first kind. The eigenequations take
the forms of $H\circ \left(
\begin{array}{c}
\overline{\Phi }_{1} \\
\overline{\Phi }_{2}%
\end{array}%
\right) =\mu \left(
\begin{array}{c}
\overline{\Phi }_{1} \\
\overline{\Phi }_{2}%
\end{array}%
\right) $. For the nonlinear system, the number of the eigenstates may be
larger than the dimension of the Hilbert space and the eigenstate could be
unstable\cite{liu}. We have obtained four eigenstates for the case $Z>$ $%
\rho $ by solving the above eigenequation. Three of them are stable and one
is unstable. We choose following stable eigenstate for example to illustrate
our theory, i.e.,$\overline{\Phi }_{1} =\left[ \frac{1}{2}\left( 1-\sqrt{1-%
\frac{\rho ^{2}}{Z^{2}}}\right) \right] ^{1/2},\text{ } $$\text{\ }\overline{%
\Phi }_{2} =\left[ \frac{1}{2}\left( 1+\sqrt{1-\frac{\rho ^{2}}{Z^{2}}}%
\right) \right] ^{1/2}e^{i\varphi },$ with eigenvalue of $\mu =Z.$

We choose the overall phase $\lambda =-\arg \left( \Psi _{1}\right) ,$ from
GP equation (\ref{sss}), we have%
\begin{equation}
\frac{d\lambda }{dt}=-i\langle (\Phi _{1},\Phi _{2})|\frac{\partial }{%
\partial t}|(\Phi _{1},\Phi _{2})\rangle +E(\Phi _{1},\Phi _{2})+\frac{Z}{2}%
(|\Phi _{1}|^{4}+|\Phi _{2}|^{4}).\text{ }
\end{equation}%
In the adiabatic evolution, we assume $\Phi _{i}=\overline{\Phi }%
_{i}(\varphi )+\delta \Phi _{i}(\varphi ),$ where $\delta \Phi _{i}$ is the
fluctuation depending on the adiabatic parameter of order $\varepsilon
=d\varphi /dt.$ Then we have the explicit expression of the zeroth order,
\begin{equation}
\alpha _{0}(\varepsilon ^{0})=Z,
\end{equation}%
and using conservation of the particle $|\Phi _{1}|^{2}+|\Phi _{2}|^{2}=1,$%
and the fact $\arg \Phi _{1}=0,$we have the first-order term,
\begin{eqnarray}
&&\alpha _{1}(\varepsilon ^{1})=-i\overline{\Phi }_{2}\frac{\partial }{%
\partial t}\overline{\Phi }_{2}  \notag \\
&&+Z\left\langle (2|\overline{\Phi }_{2}|^{2}-1)\overline{\Phi }_{2}^{\ast
},(2|\overline{\Phi }_{2}|^{2}-1)\overline{\Phi }_{2}|(\delta \overline{\Phi
}_{2},\delta \overline{\Phi }_{2}^{\ast })\right\rangle .\text{ }
\end{eqnarray}

Berry term of the geometric phase is readily deduced,
\begin{equation}
\gamma _{B}\mathbf{=}\pi \left( 1+\sqrt{1-\frac{\rho ^{2}}{Z^{2}}}\right) .
\label{gbb}
\end{equation}

Now we are going to derive the additional term which gives the $\gamma _{NL}$
of the geometric phase. Let us introduce the new variables $(n,\theta )$
through $(\Phi _{1},\Phi _{2})=(\sqrt{1-n},\sqrt{n}e^{i\theta }).$
Substitute $(\Psi _{1},\Psi _{2})=e^{-i\int^{t}\beta dt}(\Phi _{1},\Phi _{2})
$ into Eq. (\ref{sss}), and separate the real and imaginary parts, we have
four differential equations, in which two of them are identical due to the
norm conversation,
\begin{eqnarray}
\frac{d}{dt}n &=&-\rho \sqrt{n-n^{2}}\sin (\theta -\varphi ),\text{ }
\label{a1} \\
\frac{d}{dt}\theta  &=&-\frac{\rho \sqrt{1-n}}{2\sqrt{n}}\cos (\theta
-\varphi )-Zn+\beta ,  \label{a2} \\
\beta  &=&Z\left( 1-n\right) +\frac{\rho }{2}\sqrt{\frac{n}{1-n}}\cos
(\theta -\varphi ).  \label{a3}
\end{eqnarray}%
The eigenstate make up the fixed point of equations (\ref{a1}) and (\ref{a2}%
), i.e., $\overline{n}=\frac{1}{2}\left( 1+\sqrt{1-\frac{\rho ^{2}}{Z^{2}}}%
\right) $,$\overline{\theta }=\varphi .$ Let us make perturbation expansion
around the eigenstate with $n=\overline{n}(\varphi )+\delta n,\theta =%
\overline{\theta }(\varphi )+\delta \theta .$ Then inserting the above
expansion into the equations (\ref{a1}) and (\ref{a2}), and ignoring the
higher order terms of $\frac{\partial \delta n}{\partial t},\frac{\partial
\delta \theta }{\partial t},$ we obtain that%
\begin{equation}
\left(
\begin{array}{c}
\frac{\partial \overline{n}}{\partial \varphi } \\
\frac{\partial \overline{\theta }}{\partial \varphi }%
\end{array}%
\right) \frac{d\varphi }{dt}=\mathcal{L}\left(
\begin{array}{c}
\delta n \\
\delta \theta
\end{array}%
\right) ,
\end{equation}%
with
\begin{equation}
\mathcal{L}=\left(
\begin{array}{cc}
0 & -\rho \sqrt{\overline{n}-\overline{n}^{2}} \\
-2Z+\frac{\rho }{4(\overline{n}-\overline{n}^{2})^{3/2}} & 0%
\end{array}%
\right) .
\end{equation}%
Then, we have
\begin{equation}
\gamma _{NL}=\int_{0}^{2\pi }\left\langle \Lambda |\Pi \mathcal{L}^{-1}\circ
\left( \frac{\partial \overline{n}}{\partial \varphi },\frac{\partial
\overline{\theta }}{\partial \varphi }\right) ^{T}\right\rangle d\varphi .
\label{lfa}
\end{equation}%
where $\Lambda =Z((2\overline{n}-1)\sqrt{\overline{n}}^{-i\overline{\theta }%
},(2\overline{n}-1)\sqrt{\overline{n}}^{i\overline{\theta }})$. Finally, we
get
\begin{equation}
\gamma _{NL}=\frac{\pi \rho ^{2}}{Z\sqrt{Z^{2}-\rho ^{2}}}.  \label{bb}
\end{equation}%
%

The above new form of geometric phase is verified numerically by directly
integrating the Schr\"odinger Eq.(19) using Runge-Kutta algorithm.

In summary, we show that the Berry phase associated with the adiabatic
evolution of BECs has been modified by the nonlinearity. This nonlinear
correction is of significance because it could affect the interferences of
the matter waves. It is expected to be observed in future's experiments. Our
theories, on the other aspect, cover the general nonlinear systems in
nonlinear optics and condense matter physics.
Since some of the theories covered by our results are meanfield limits of
quantum many-body theories, the possibility of generalizing these
considerations to quantized field theories from the correspondence principle
is of great interest for future study.

This work is supported by National Natural Science Foundation of China
(No.10725521,10604009), 973 project of China under Grant No. 2006CB921400,
2007CB814800. 


\begin{thebibliography}{99}
\bibitem{theorem1} P. Ehrenfest, Ann. Phys. (Berlin) 51, 327 (1916); M. Born
and V. Fock, Z. Phys. 51, 165 (1928); L. D. Landau, Zeitschrift 2, 46
(1932); C. Zener, Proc. R. Soc. A 137, 696 (1932); J. Schwinger, Phys. Rev.
51, 648 (1937).

\bibitem{theorem2} M.V. Berry, Proc. R. Soc. Lond. A \textbf{392}, 45 (1984).

\bibitem{liu2} \emph{Geometric Phase in Physics}, edited by A. Shapere and
F. Wilczek (World Scientific, 1989); \emph{The Geometric Phase in Quantum
Systems}, A. Bohm, A. Mostafazadeh, H.Koizumi, Q.Niu, J.Zwanziger,
(Springer, 2003).

\bibitem{application} K. Bergmann, H. Theuer, and B.W. Shore, Rev. Mod.
Phys. 70, 1003 (1998).

\bibitem{fu3} F. Wilczek and A. Zee, Phys. Rev. Lett. 52 2111 (1984).

\bibitem{fu4} Y. Aharonov and J.S. Anandan, Phys. Rev. Lett. 58 1593(1987).


\bibitem{renew} Quantum Computation and Quantum Information Theory, edited
by C. Macchiavello, G.M. Palma, and A. Zeilinger (World Scientific,
Singapore, 2000); J. Jones,V.Vedral, A. K. Ekert, and C. Castagnoli, Nature
(London) 403, 869 (2000); G. Falci, R. Fazio, G.M. Palma, J. Siewert, and
V.Vedral, Nature (London) 407, 355 (2000); L.-M. Duan, I. Cirac, and P.
Zoller, Science 292, 1695 (2001).

\bibitem{fu6} Chuanwei Zhang, Artem M. Dudarev, and Qian Niu, Phys. Rev.
Lett. \textbf{97}, 040401 (2006);Di Xiao, Yugui Yao, Zhong Fang, and Qian
Niu,Phys. Rev. Lett. \textbf{97}, 026603 (2006); Wang Yao and Qian Niu,
Phys. Rev. Lett. \textbf{101}, 106401 (2008).

\bibitem{bec}
C. J. Pethick, H. Smith, 'Bose-Einstein Condensation in Dilute Gases',
Cambridge University Press (2002); 'Bose-Einstein Condensation in Atomic
Gases', edited by M. Inguscio, S. Stringari, and C. E. Wieman (IOS Press,
Amsterdam), 1999.


\bibitem{liu} Jie Liu, Biao Wu, and Qian Niu, Phys. Rev. Lett. \textbf{90},
170404 (2003).

\bibitem{pu} Han Pu, Peter Maenner, Weiping Zhang, and Hong Y. Ling, Phys.
Rev. Lett. \textbf{98}, 050406 (2007).


\bibitem{chiao} For example, see, J. C. Garrison, R.Y. Chiao, Phys. Rev.
Lett. \textbf{60}, 165 (1988); Biao Wu, Jie Liu, and Qian Niu, Phys. Rev.
Lett. 94, 140402 (2005); F N Litvinets et al J. Phys. A: Math. Gen. \textbf{%
39} 1191 (2006); X. X. Yi, X. L. Huang, and W. Wang, Phys. Rev. A \textbf{77}%
, 052115 (2008).

\bibitem{GP} E. P. Gross, Nuovo Cimento \textbf{20}, 454 (1961); L. P.
Pitaevskii, Zh. Eksp. Teor. Fiz. \textbf{40}, 646 (1961); [Sov. Phys. JETP
\textbf{13}, 451 (1961)].

\bibitem{twomode} A. Smerzi, S. Fantoni, S. Giovanazzi, and S. R. Shenoy,
Phys. Rev. Lett. 79, 4950 (1997); A. J. Leggett, Rev. Mod. Phys. 73, 307
(2001).


\end{thebibliography}
\end{document}